\documentclass[prd,twocolumn,
nofootinbib,preprintnumbers]{revtex4}

\usepackage{times}
\usepackage{amssymb,amsbsy,amsmath,amsfonts}
\usepackage{mathrsfs}
\usepackage{graphicx}
\usepackage{graphics,psfrag,longtable}
\usepackage{nicefrac}
\usepackage{esint}
\usepackage{upgreek}

\usepackage[colorlinks,linktoc=all,citecolor=blue,linkcolor=red]{hyperref} 

\DeclareMathAlphabet{\mathantt}{OML}{antt}{l}{it}
\DeclareMathAlphabet{\mathpzc}{OT1}{pzc}{m}{n}

\def\beq{\begin{equation}}
\def\eeq{\end{equation}}
\def\bea{\begin{eqnarray}}
\def\eea{\end{eqnarray}}
\def\beqa{\begin{equation}\begin{array}{l}}
\def\eeqa{\end{array}\end{equation}}
\def\eqlab#1{\label{eq:#1}}

\def\eref#1{(\ref{eq:#1})}
\def\Eqref#1{Eq.~(\ref{eq:#1})}

\def\half{\mbox{$\frac{1}{2}$}}

\def\third{\mbox{$\frac{1}{3}$}}
\def\sixth{\mbox{$\frac{1}{6}$}}

\def\barr{\left(\begin{array}{c}}
\def\earr{\end{array}\right)}
\def\bmat{\left(\begin{array}{cc}}
\def\emat{\end{array}\right)}
\def\al{\alpha}

 \def\De{\Delta}
  \def\eps{\epsilon}

\def\nn{\nonumber}
\def\dd{\mathrm{d}}

\DeclareMathOperator\im{Im}
\def\3d{3-D}

\def\ol#1{\overline{#1}}

\begin{document}
\preprint{MITP/15-009}
\title {Breakdown of the expansion of finite-size corrections to the hydrogen Lamb shift
 in moments of charge distribution}

\author{Franziska Hagelstein}
\author{Vladimir Pascalutsa}
\affiliation{
Institut f\"ur Kernphysik, Cluster of Excellence PRISMA,  Johannes Gutenberg-Universit\"at Mainz, D-55128 Mainz, Germany}

\begin{abstract}
We quantify a limitation in the usual accounting of the finite-size effects, where the leading $[(Z\alpha)^4]$ and subleading  $[(Z\alpha)^5]$ contributions to the Lamb shift are given by
the mean-square radius and the third Zemach moment of the charge distribution. In the presence of any
non-smooth behaviour of  the nuclear form factor at scales comparable to the inverse Bohr radius, the expansion of the 
Lamb shift in the moments breaks down. This is relevant
for some of the explanations of the ``proton size puzzle''. We find, for instance,
that the de R\'ujula toy model of the proton form factor does not resolve the puzzle as claimed, despite the large value of the third Zemach moment. Without relying on the radii expansion,
we show how tiny, milli-percent (pcm) changes in the proton electric form factor at a MeV scale would be able to explain the puzzle. It shows that one needs to know all the soft contributions to proton
electric form factor to pcm accuracy for a precision extraction of the proton charge radius from
atomic Lamb shifts.
\end{abstract}
\date{\today}
\maketitle

\section{Introduction}
The proton structure is long-known to affect the hydrogen spectrum,
predominantly by an upward shift of the $S$-levels expressed
in terms of the root-mean-square (rms) radius, 
\beq
\eqlab{moments}
R_E =  \sqrt{\langle r^2\rangle_E}, \quad \langle r^N\rangle_E
\equiv \int \! \dd \vec r\,\,  r^N \! \rho_{E}(\vec r),
\eeq  
of the proton charge distribution $\rho_{E}$. At leading order (LO)
in the fine-structure constant $\al$, the $n$th $S$-level is shifted by (cf., \cite{Eides:2000xc}):
\beq
\eqlab{LO}
\De E_{nS}(\mbox{LO}) = \frac{2(Z\al)^4 m_r^3}{3n^3} R_E^2,
\eeq
where $Z=1$ for the proton, $m_r$ is the reduced mass. The proton
charge radius has thus been extracted from the hydrogen ($e$H)
and muonic-hydrogen ($\mu$H) Lamb shifts, with rather contradictory
results:
\begin{subequations}
\bea
R_{Ep}(e\mathrm{H}) &=& 0.8758(77) \, \mbox{fm \cite{Mohr:2012tt}}
\eqlab{reH}
, \\
R_{Ep} (\mu\mathrm{H}) & = & 0.84087(39) \, \mbox{fm \cite{Antognini:1900ns,Antognini:2013rsa}}.
\eqlab{rmuH}
\eea 
\end{subequations}
The $e$H value  is backed up by the extractions from
electron-proton ($ep$) scattering~\cite{Bernauer:2010wm,Zhan:2011ji}, albeit
with a notable exception~\cite{Lorenz:2014yda}.

The next-to-leading order (NLO) effect of the nuclear charge distribution is
given by \cite{Friar:1978wv}:
\beq
\eqlab{NLO}
\De E_{nS}(\mbox{NLO}) = -\frac{(Z\al)^5 m_r^4}{3n^3}  R_{E(2)}^3, 
\eeq
with $R_{E(2)} =\sqrt[3]{\langle r^3\rangle_{E(2)} }$
the Friar radius and 
\beq
\langle r^3\rangle_{E(2)} = \int \! \dd \vec r\, \rho_{E}(\vec r\,) \int \!\dd \vec r\, ' \, | \vec r -\vec r\, ' |^3 \, \rho_{E}(\vec r \, ')
\eeq
the third Zemach moment. Other $\al^5$ effects of proton structure, such as 
polarizabilities, play a lesser role in both normal and muonic hydrogen, 
and are not in anyway of relevance to the present discussion of finite-size effects. 

A Lorentz-invariant definition of the above moments is given in terms of the
electric form factor (FF),  $G_E(Q^2)$, as:
\begin{subequations}
\bea
\langle r^2\rangle_E &=& -6 \frac{\dd}{\dd Q^2} G_E(Q^2)\Big|_{Q^2= 0}
,\label{r2}\\
\langle r^3\rangle_{E(2)} &=& \frac{48}{\pi} \int_0^\infty \!\frac{\dd Q}{Q^4}\,
\left\{ G_E^2(Q^2) -1 +\third  Q^2\langle r^2\rangle_E\right\}. \qquad 
\eea
\end{subequations}

At the current level of precision, the $e$H Lamb shift sees only the LO term, while in $\mu$H
the NLO term becomes appreciable. An immediate resolution of the $e$H vs.\
$\mu$H discrepancy (aka, the {\it proton size puzzle}) was suggested by de R\'ujula \cite{DeRujula:2010dp}, whose
toy model for proton charge distribution yielded a large Friar radius,
capable of providing the observed $\mu $H Lamb shift using the $R_E$ value from $e$H.
Shortly after, this model was shown to be incompatible with the empirical FF $G_E$ extracted
from $ep$ scattering \cite{Miller2011,Distler:2010zq}. In this work we find that the $\mu$H Lamb shift in de R\'ujula's model
is not described correctly by the standard formulae of Eqs.~\eref{LO} and \eref{NLO}. The correct result involves
an infinite series of moments, and  it does not provide any significant reduction of the discrepancy in that model. 
We shall consider a different scenario of mending the discrepancy by  a small change in the proton FF, using the corrected
formulae. 

\section{Lamb shift: to expand or not}
Our main observation is that the standard expansion in the moments is only valid provided
the convergence radius of the Taylor expansion of $G_E$ in $Q^2$ is much larger than
the inverse Bohr radius of the given hydrogen-like system. In other words,
for $Q^2\sim (Z\al m_r)^2$,  the electric FF must be representable
by a quickly convergent power series. 

\begin{figure}[t] 
    \centering 
       \includegraphics[scale=0.45]{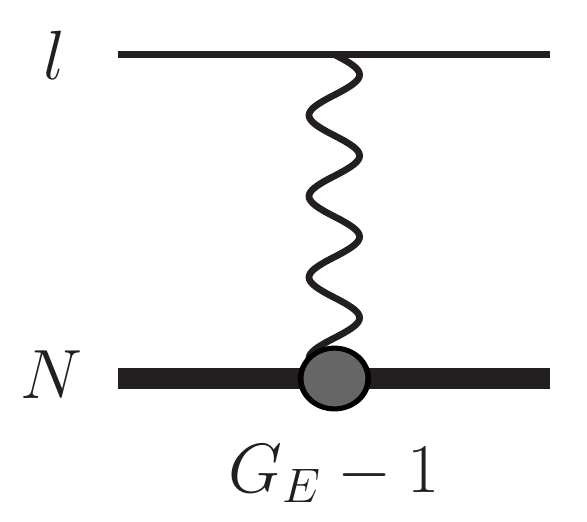}
       \caption{One-photon exchange
graph with FF dependent electromagnetic vertex.}       
       \label{fig:1gamma}
\end{figure}

To see this we write the electric FF 
correction to the Coulomb potential ($-Z\al/r$) as follows:
\beq
V_{\mathrm{FF}} (r) =\frac{Z\al}{\pi r} \int_{t_0}^\infty \! \frac{\dd t}{t} \, e^{-r \sqrt{t}} \, \im G_E(t),
\eeq
where $\im G_E$ is the discontinuity in the FF across the branch cuts in the time-like
region. This potential is derived by taking the non-relativistic limit of the one-photon exchange
graph in Fig.\ref{fig:1gamma}, and assuming the FF to satisfy the subtracted dispersion relation:
\beq
\eqlab{DRel}
G_E(Q^2) = 1- \frac{Q^2}{\pi} \int^\infty_{t_0} \! \! \frac{\dd t}{t}\, \frac{\im G_E(t)}{t+Q^2}\,.
\eeq
Here, as usual: $Q^2=-q^2 \geq 0$, with $q$ the four-momentum of the exchanged
photon; $t_0 \geq 0$ is the particle-production threshold associated with the given discontinuity of the FF.
The use of the dispersion relation here is very much analogous to the Schwinger's method of calculating
the vacuum polarization (Uehling) effect \cite{Schwinger1949}.

The effect of this Yukawa-type correction on the hydrogen spectrum can easily be 
worked out  using time-independent perturbation theory. At first order we obtain the
 following expression for the $2P$-$2S$ Lamb shift:
\begin{subequations}
 \bea
 E^{\mathrm{FF}(1)}_{2P-2S} & =& -\frac{(Z\al)^4 m_r^3}{2\pi} \int_{t_0}^\infty \!\!\dd t \, \frac{\im G_E (t)}{(\sqrt{t}+Z\al m_r)^4} 
\eqlab{rmsLSa}\\
&=& -\frac{(Z\al)^4 m_r^3}{12} \sum_{k=0}^\infty 
 \frac{(-Z\al m_r)^{k}}{k!} \langle r^{k+2}\rangle_E,\quad
\eea
\end{subequations}
where in the second step we have expanded in the moments of the charge distribution, introduced in \Eqref{moments}, using the following (Lorentz-invariant) definition:
\beq
\eqlab{rmsdef}
\langle r^N\rangle_E =
\frac{(N+1)!}{\pi}\int_{t_0}^\infty\!\! \dd t \, \frac{\im  G_E(t)}{t^{N/2+1 }  }.
\eeq

One can now see that the convergence radius of the power-series expansion in moments is limited
by $t_0$,  i.e.\ the proximity of the nearest particle-production threshold. 
For the proton we of course expect $t_0$ to be a hadronic scale, of which the pion mass is the lowest, and the
series should converge quickly for hydrogen, and  in fact for most of the hydrogen-like systems.
The new physics search is a different matter.  Inclusion of new light particles,
or anything that changes the FF at very low $Q^2$ (and equivalently, the charge distribution at large distances) may invalidate the expansion in the moments. Before illustrating this point more quantitatively (in Sect.~\ref{sec:toy}), 
we need to make a few more  technical steps.

First, we observe that the expansion of the first-order Lamb shift to order $\al^5$, 
\bea
 E^{\mathrm{FF}(1)}_{2P-2S}  &=& -\frac{(Z\al)^4 m_r^3}{12} \left[ \langle r^2\rangle_E - Z\al m_r 
 \langle r^3\rangle_E \right], 
 \label{LS1Exp}
 \eea
 with
 \beq
 \langle r^3\rangle_{E} = \frac{48}{\pi} \int_0^\infty \!\frac{\dd Q}{ Q^{4}}\,
\left\{ G_E(Q^2) -1 +\frac16 Q^2 \langle r^2\rangle_E \right\},\label{r3}
\eeq 
 does not exactly reproduce the Friar term in \Eqref{NLO}. To do so, we need
 to treat the potential to second order in perturbation theory. The second-order
 Lamb-shift is, to $O(\al^5)$, given by: 
 \begin{subequations}
 \bea
E^{\mathrm{FF}(2)}_{2P-2S} &=&  (Z\al)^5 m_r^4 \,\frac{2}{\pi}
\int_0^\infty \!\! \dd Q\, 
\bigg\{\frac{1 }{\pi}  \int_{t_0}^\infty \!\frac{\dd t}{t} \,
\frac{ \im G_E(t) }{t+Q^2} \bigg\}^2  \nn\\
&=&  (Z\al)^5 m_r^4  \,
\frac{2}{\pi}\int_0^\infty \!\frac{\dd Q}{Q^4}\, \left\{ G_E(Q^2) -1 \right\}^2\\
& =& -\frac{(Z\al)^5 m_r^4 }{12} \Big[
\langle r^3\rangle_E - \half \langle r^3\rangle_{E(2)} \Big].
 \eea
 \end{subequations}
Combining it with the first order, we see that the third moment is replaced by the 
third Zemach moment, leading to \Eqref{NLO}. Certainly, if the expansion of the first-order result
in moments breaks down, this second-order calculation is inadequate too. In either case, however,
the second-order shift is suppressed by $Z\al$, with respect to the the first-order shift.

It is furthermore convenient to express the exact (unexpanded) first-order Lamb shift in terms of $G_E$, rather than 
of its discontinuity.  Observing  that in any identity of the form,  
\beq
\frac{1}{\pi } \int_{t_0}^\infty\!\dd t\, W(t) \im G_E(t) = \int_0^\infty \!\dd Q\,w(Q) \, G_E(Q^2),
\eeq
the function $W$ is the Stieltjes integral transform of $w$, i.e.:
\beq
W(t) = \int_0^\infty \!\! \dd Q \, \frac{w(Q)}{t+Q^2}.
\eeq
we  recast \Eqref{rmsLSa} in terms of $G_E$ by computing the inverse Stieltjes transform \cite{Schwarz2005} of
\beq
W(t) =  -\frac{(Z\al)^4 m_r^3}{2(\sqrt{t}+\al m_r)^4}, 
\eeq
that is:
\bea
\eqlab{iST}
w(Q)&=&  \frac{Q}{i\pi} \lim_{\varepsilon\to 0} \left\{W(-Q^2-i\varepsilon)-W(-Q^2+i\varepsilon)\right\},\nn\\
&=& - \frac4\pi  (Z\al)^5 m_r^4  \, Q^2 
\frac{ (Z\al m_r)^2-Q^2}{\left[(Z\al m_r)^2+Q^2\right]^4}.
\eea

\begin{figure}[t] 
    \centering 
       \includegraphics[scale=0.48]{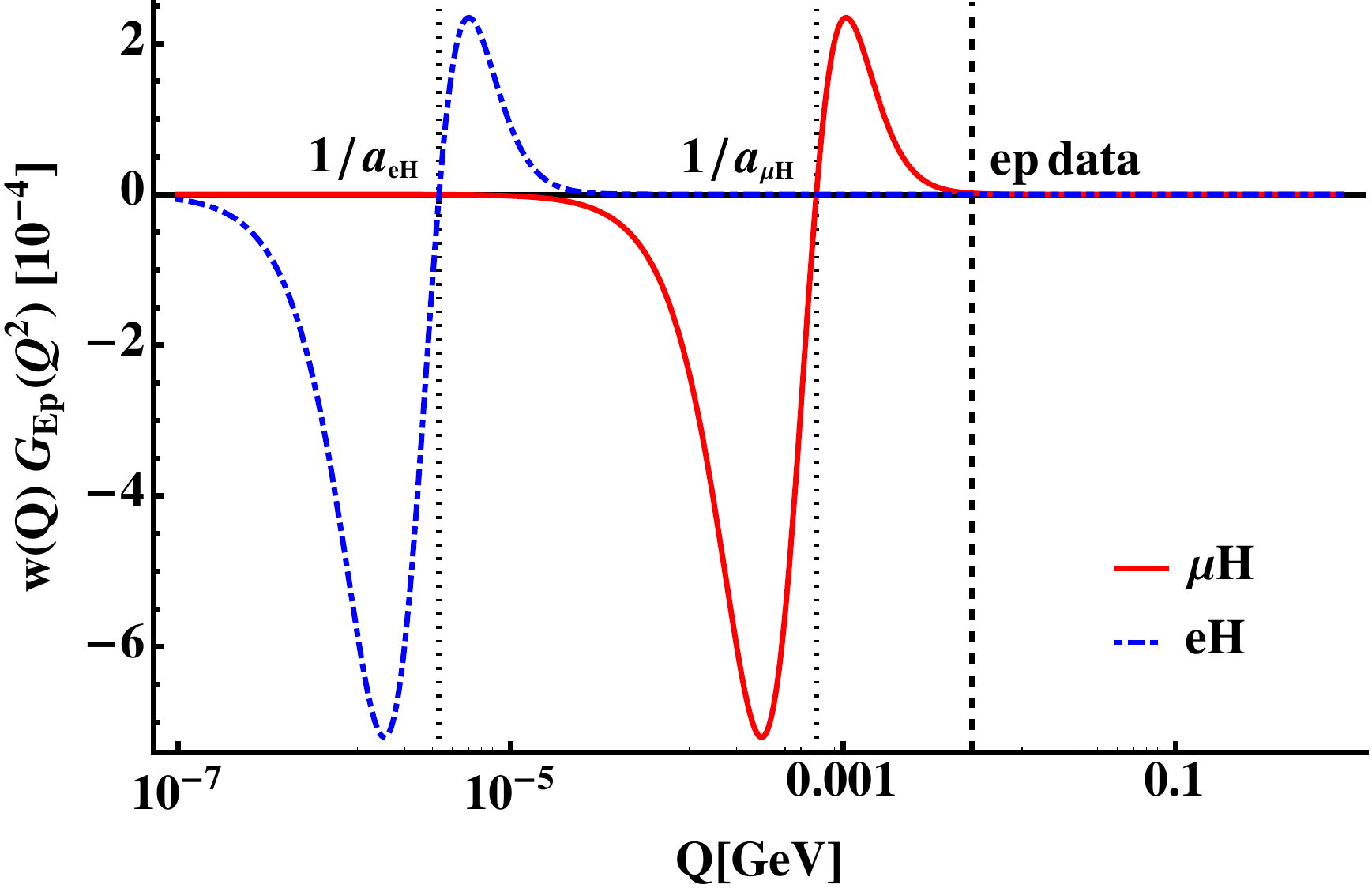}
       \caption{Integrand of the 1st-order contribution to the Lamb shift [cf.\ Eq.~(\ref{wG}) with \Eqref{iST}]
       of $e$H (blue dash-dotted line) and of $\mu$H (red solid line) for the dipole FF. The dotted vertical lines
       indicate the inverse Bohr radius of the two hydrogens, while the vertical dashed line indicates the onset of the
       $ep$ data.}       
       \label{fig:Integrand}
\end{figure}

 We also can express the Lamb shift in terms of the spherically-symmetric
 charge distribution, knowing that the latter is given by the following Laplace transform of the FF discontinuity:
 \beq
 \rho_E (r) = \frac{1}{(2\pi)^2\, r} \int^\infty_{t_0} \! \! \dd t\, \im G_E(t)\,  e^{-r\sqrt{t}} .
 \eeq
 Collecting it all together, we have
 \begin{subequations}
 \eqlab{wGall}
   \bea
   \label{wG}
 E^{\mathrm{FF}(1)}_{2P-2S} & =& \int_0^\infty \!\dd Q\,w(Q) \, G_E(Q^2)\\
&=& -\third \pi (Z\al)^4 m_r^3  \int_0^\infty\! \dd r \, r^4 e^{-r/a } \rho_E (r),\quad
 \eea
 \end{subequations}
 with $w(Q)$ given in \Eqref{iST}, and the Bohr radius $a=1/(Z\al m_r)$. 
 The expression in terms of the charge density is the simplest and has the most intuitive
 interpretation: the first-order Lamb shift is given by the mean-square radius cut off at the Bohr radius by the Coulomb wave function. Indeed, it is simply the LO result, \Eqref{LO}, with $\langle r^2 \rangle_E$ replaced by 
 $\langle r^2 e^{-r/a }  \rangle_E$.
 
The expression in terms of the FF, Eq.~(\ref{wG}), can be derived without the reference to the dispersion relation.
Instead, one can follow the momentum-space derivation of the  Wichmann-Kroll  contribution 
in Ref.~\cite{Karshenboim:2010cq}. The weighting function $w(Q)$ then arises as a convolution 
of the hydrogen wave functions, which is interpreted as the atomic FF. The use of the dispersion relation
for the proton form factor will, however, allow for a simple and systematic account of the retardation effects,
as will be discussed elsewhere.
 
 We are now in position to plug in a specific charge density or a FF parametrisation and see how good the expansion
 in the moments is. For instance, we find that the expansion does not converge 
 for the charge density proposed by de R\'ujula to explain the proton size puzzle. His model
 does give an unusually large Friar radius, however it does not affect the $\mu$H Lamb shift in the way 
 expected from the expansion into moments. We find in this model that, in the corner of the parameter space
 used to explain the puzzle,  the $\mu$H Lamb shift obtained using \Eqref{wGall} is very similar to the LO result, \Eqref{LO}, and is nearly not as sensitive to the  Friar-radius value as the NLO term, \Eqref{NLO}, by itself.
 In contrast, for most of the popular FF parametrisations used in fitting the $ep$ data, including
 the dipole form,  the expansion in moments is appropriate. We have checked all of the parametrisation listed
 in Appendix of Ref.~\cite{Karshenboim2014}.
 
 In general, the breakdown of the moment expansion can only be expected for FF with 
 non-smooth behaviour in the region of $Q$ comparable to inverse Bohr radius. 
 To see this more explicitly we plot the integrand of Eq.~(\ref{wG}) in Fig.~\ref{fig:Integrand}, for the dipole form of the
 proton FF, i.e.: $G_{Ep}= (1+Q^2/0.71\mathrm{GeV}^2)^{-2}$. One can see that the total integral
 over $Q$ is the result of large cancellations around the Bohr radius scale. Any relatively small
 variation in the FF around that scale may lead to significant effects in the Lamb shift. 
In order to define a proper charge radius which could be determined from both the atomic and
scattering experiments it is therefore mandatory to decompose the FF into the ``smooth" and
 ``non-smooth'' parts. The contribution of the former can then be expanded in moments,
while the  latter must be treated exactly.  This is already done in the case of QED contributions to the FF.
 
 Concerning the proton size puzzle, let us note that the two curves in Fig.~\ref{fig:Integrand}
 represent the two hydrogens, and that their regions of large cancellations are well separated.
 Also separated is the region of the existing  $ep$ scattering data, $ Q^2 > 0.004\, \mbox{GeV}^2$. 
 Obviously, if there is a small missing effect in the FF  responsible for the puzzle, 
it must be localised near one of the two inverse  Bohr radii, where its impact is maximised. 
It also should not contradict the $ep$ data, hence the allowed region is
anywhere from 0 to roughly 50 MeV. 
On the other hand, only in $\mu$H the radius is extracted from the classic
Lamb shift, the extraction from H combines a number of different transitions. Hence, 
the simplest way to solve the puzzle is to have the missing effect in the $\mu$H Lamb shift, i.e.,
to place it between 1 and 50 MeV. 
 To see this more quantitatively, we next consider a toy model for the possible effect in
 the FF in that region.
 
 \section{Resolving the puzzle}
 \label{sec:toy}
 
 We assume the electric FF to separate into a smooth ($\ol G_E$) and non-smooth part ($\widetilde G_E$), such that,
 \beq
  G_E(Q^2) = \ol G_E(Q^2) + \widetilde G_E(Q^2).
  \eeq 
 For the smooth part we shall take a well-known parametrisation which fits the $ep$ data, while
 for the non-smooth one we consider a single Lorentzian:\footnote{This anzatz turned out to be similar in spirit to the one considered by Wu and Kao \cite{WuKao}, who, however, deduced its effect on the Lamb shift through the expansion in moments, thus leading to incorrect results.}
 \beq
\widetilde G_E(Q^2)=\frac{A}{\pi}\left\{\frac{\epsilon^2}{\big(Q^2-Q_0^2\big)^2+\epsilon^4}-\frac{\epsilon^2}{Q_0^4+\epsilon^4}\right\},\label{FFcorr}
\eeq 
which describes a Breit-Wigner type of peak around $Q_0^2$ with the width given by $2\eps^2$, and which does not
contribute to the charge: $\widetilde G_E(0)=0 $.

According to Fig.~\ref{fig:Integrand}, 
in order to make a maximal impact on the puzzle, the fluctuation $\widetilde G_E$ must be located at the extremi [of $w(Q)$ in Eq.~(\ref{wG})] around of either
$e$H or $\mu$H inverse Bohr radius. Here we shall only consider the latter case
and set:
\beq 
Q_0 = \frac{1}{2a_{\mu H}}\sqrt{5+\sqrt{17}} \simeq 1.04657 \, \mbox{MeV}. 
\eqlab{Q0}
\eeq
This means we place a small fluctuation exactly on top of the maximum above $1/a_{\mu H}$ in Fig.~\ref{fig:Integrand}.
The sensitivity of our results to the precise value of $Q_0$ is very mild for $Q_0$ between 1 and 10 MeV. 

This choice of $Q_0 > 1/a_{\mu H}$ conditions the choice of the smooth part, in case one wants to solve the puzzle.
Indeed, since the non-smooth part affects mostly the $\mu$H result, 
the smooth part must have a radius consistent with the $e$H value.
We therefore adopt the chain-fraction fit of Arrington and Sick \cite{Arrington:2006hm}:
\beq
\ol G_E(Q^2)=\frac{1}{1+\frac{3.478 \,Q^2}{1-\frac{0.140 \,Q^2}{1-\frac{1.311 \,Q^2}{1+\frac{1.128 \,Q^2}{1-0.233 \,Q^2}}}}} 
\eeq

The other two parameters parametrising $\widetilde G_E$,
 $A$ and $\eps$, are fixed by requiring our FF to yield the empirical 
Lamb shift contribution, in both normal and muonic hydrogen, i.e.:  
 \begin{subequations}
 \eqlab{LS}
 \bea
&& E^{\mathrm{FF}(exp.)}_{2P-2S}(e\mathrm{H}) = -0.620(11) \, \mbox{neV}, \label{LSeH}\\
&& E^{\mathrm{FF}(exp.)}_{2P-2S} (\mu\mathrm{H})  = -3650(2) \,  \mbox{$\upmu$eV} \label{LSmuH}.
\eea 
 \end{subequations}
Note that these are not the experimental Lamb shifts, but only the finite-size contributions, described by Eqs.~\eref{LO}, \eref{NLO}, with the corresponding empirical values for the radii. In the $e$H case we have taken the CODATA value of the proton radius, \Eqref{reH}, which is obtained as an weighted average over several hydrogen spectroscopy measurements, and $R_{E(2)} = 2.78(14)$ fm \cite{Borie:2012zz}.  In the
 $\mu$H case we have taken the values from Ref.~\cite{Antognini:2013rsa}, hence
 \Eqref{rmuH} for the radius and the same as the above value for $R_{E(2)}$.

  \begin{figure}[tb] 
    \centering 
       \includegraphics[scale=0.48]{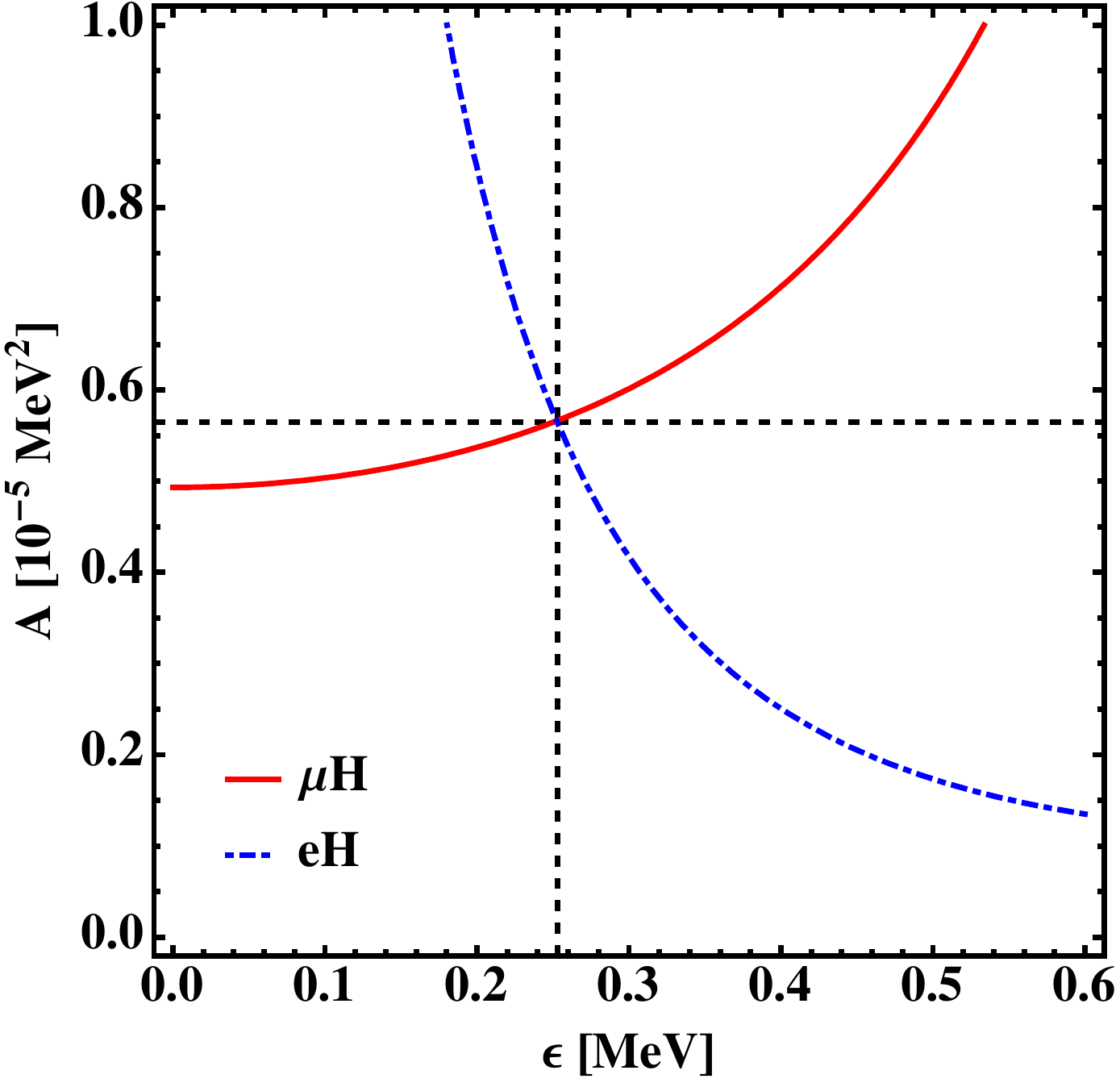}
       \caption{Parameters of $\widetilde G_E$ for which the $e$H and $\mu$H Lamb shifts of \Eqref{LS} 
       are reproduced, given $Q_0$ set by \Eqref{Q0}. }
              \label{fig:Parameter}
\end{figure}

  Figure~\ref{fig:Parameter} shows at which $A$ and $\eps$ our
 FF complies with either $e$H (dot-dashed curve) or $\mu$H (solid curve) Lamb shift. For
 $A=5.65\times10^{-6}$ MeV$^2$ and $\epsilon=0.253$ MeV, our FF describes them both, thus solving the puzzle.
 
 We emphasise that the magnitude of the fluctuation in the FF is extremely tiny, 
\beq
\big\vert\widetilde G_E/\, \ol G_E\big\vert <3 \times 10^{-5} ,
\eeq
for any $Q^2$. Nevertheless, it obviously has a profound effect on $\mu$H Lamb shift.
 Its effect on the second and third moments is given by:
\bea
\widetilde{\langle r^2\rangle}_E&\equiv &-6 \frac{\dd}{\dd Q^2} \widetilde G_E(Q^2)\Big|_{Q^2= 0}
=-\frac{ 12 A  Q_0^2 \epsilon ^2}{ \pi (Q_0^4+\epsilon^4)^2},\\
\widetilde{\langle r^3\rangle}_E&\equiv &\frac{48}{\pi} \int_0^\infty \!\frac{\dd Q}{Q^4}\,
\left\{ \widetilde G_E(Q^2) +\sixth \widetilde{\langle r^2\rangle}_E Q^2\right\}\nn\\
&=&-\frac{12iA}{\pi}\left\{\frac{1}{(-Q_0^2-i\epsilon^2)^{5/2}} -\frac{1}{(-Q_0^2+i\epsilon^2)^{5/2}} \right\}\nn\\
&=&\frac{24A}{\pi Q_0^5}+O(\epsilon^2/Q_0^2).
\eea
The numerical values of these moments, together with their ``would be" effect on the Lamb shift and the non-expanded
Lamb result,  are given  
in Table \ref{Table}. One can see that the expansion in moments breaks down for the fluctuating contribution
to $\mu$H.

\begin{table}[t]
\begin{tabular}{c|c|c|c|c}
&Eq.&$\ol G_E$&$\widetilde G_E$&$ G_E$\\
\hline
$\langle r^2\rangle_E \, [\mbox{fm}^2]$&(\ref{r2})&$(0.9014)^2$&$-(0.2016)^2$&$(0.8786)^2$\\
$\langle r^3\rangle_E \,[\mbox{fm}^3]$&(\ref{r3})&$(1.052)^3$&$(6.384)^3$&$(6.394)^3$\\
\hline
Lamb-shift, expanded & (\ref{LS1Exp}) & && \\
$E_{2P-2S}^{\mathrm{FF}(1)}(e\mathrm{H})[\text{neV}]$ & &$-0.65690$&$0.03684$& $-0.62006$\\
$E_{2P-2S}^{\mathrm{FF}(1)}(\mu\mathrm{H})[\upmu\text{eV}]$& &$-4202$&$4961$&$759$\\
\hline
Lamb-shift, exact & (\ref{wG}) & && \\
$E_{2P-2S}^{\mathrm{FF}(1)}(e\mathrm{H}) [\text{neV}]$&&$-0.65691$&$0.03684$&$-0.62007$\\
$E_{2P-2S}^{\mathrm{FF}(1)}(\mu\mathrm{H})[\upmu\text{eV}]$&&$-4202$&$551$&$-3651$\\
\end{tabular}
\caption{Lamb shift and moments corresponding to our model FF, with $\epsilon=0.253\,\text{MeV}$, $A=5.65\times10^{-6}$ MeV$^2$, and $Q_0$ set by \Eqref{Q0}.}
\label{Table}
\end{table}

We furthermore checked that by moving the bump (i.e., the Lorentzian structure) in $G_E$ to higher $Q$, anywhere between 
1 and 50 MeV, we can obtain a stable solution conforming both hydrogens. However, starting from  $Q_0 \approx 5$ MeV 
the width and the magnitude of the bump quickly increase, reaching percent effects for $Q_0 \gtrsim 20$ MeV. We find thus that the existing $ep$ data, which start at about 60 MeV, and have the precision of a few percent, rule out the existence
of such a ``solving-the-puzzle bump'' anywhere above 50 MeV. New $ep$ experiments at lower $Q$ of similar precision 
could lower this exclusion limit to about 20 MeV.

To summarise, introducing a tiny (less than 30 ppm) bump on an empirical FF used to fit the $ep$ scattering data, 
we are able to reconcile the $e$H and $\mu$H Lamb shift results. The bump is localised near the 
inverse Bohr radius of $\mu$H, and as such affects mostly the $\mu$H result. It has a small effect
on the charge radius and the $e$H Lamb shift, and certainly no effect on the fit of the present $ep$ data.  

Of course, until one finds a physical justification for such a structure at low $Q$, one should take this model
only as a word of caution against the very optimistic  view of uncertainties in the
charge radius extractions. All the contributions to the nucleon (as well as the lepton) electric FF at low $Q$ 
should be known to better than a percent accuracy before claiming the discrepancy.

\section{Conclusion}

We have shown that the effect of the nuclear charge distribution on
the hydrogenic Lamb shift is not always expandable in the moments of charge distribution ---
a very small fluctuation in the charge distribution near the Bohr radius of the atomic system
may invalidate the expansion.  The exact (non-expanded) expression for the finite-size effect 
is given by \Eqref{rmsLSa}, or equivalently \Eqref{wGall}. 

An example of charge density which invalidates the expansion of the $\mu$H Lamb shift in the moments
is provided by  the toy model of de R\'ujula \cite{DeRujula:2010dp}. Applying \Eqref{wGall} to this model shows
no significant reduction of the discrepancy between the $e$H and $\mu$H extraction of the proton 
charge radius. 

We have set up another simple model for the proton electric form factor which resolves the puzzle. Although we
do not insist that this model has anything to do with reality, it demonstrates how a tiny ($\, \approx 10^{-5}$)
effect on the proton form factor, localised at low $Q$, may invalidate the usual line of arguments leading to the discrepancy between electronic
and muonic experiments. We hope it will motivate a theoretical search of the possible low-$Q$ effects
in the form factors, as well as the new experiments called to constrain or exclude this scenario of resolving the proton size puzzle.

\section*{Acknowledgements}
We thank Chung-Wen Kao for drawing our attention to his work after the first version of this manuscript appeared.
This work was supported by the Deutsche Forschungsgemeinschaft (DFG) through the Collaborative Research Center SFB 1044 [The Low-Energy Frontier of the Standard Model], and the Graduate School DFG/GRK 1581
[Symmetry Breaking in Fundamental Interactions].

\end{document}